\RequirePackage[running]{lineno}

\documentclass[aps,prl,nofootinbib,twocolumn,showpacs,superscriptaddress,letterpaper,amsmath,amssymb,preprintnumbers]{revtex4}

\pdfoutput=1

\usepackage{graphicx}  
\usepackage{dcolumn}   
\usepackage{bm}        
\usepackage{amssymb}   
\usepackage{feynmf}
\usepackage{slashed}
\usepackage{multirow}
\usepackage{hyperref}
\def\gsim{\lower0.5ex\hbox{$\:\buildrel >\over\sim\:$}}
\def\lsim{\lower0.5ex\hbox{$\:\buildrel <\over\sim\:$}}

\newcommand{\be}{\begin{equation}}
\newcommand{\ee}{\end{equation}}
\newcommand{\bea}{\begin{eqnarray}}
\newcommand{\eea}{\end{eqnarray}}

\newcommand{\nbox}{{\,\lower0.9pt\vbox{\hrule \hbox{\vrule height 0.2 cm
\hskip 0.2 cm \vrule height 0.2 cm}\hrule}\,}}

\def\sub#1{_{\lower.25ex\hbox{$\scriptstyle#1$}}}

\def\gev{\,{\rm GeV}}

\newskip\zatskip \zatskip=0pt plus0pt minus0pt
\def\matth{\mathsurround=0pt}
\def\lsim{\mathrel{\mathpalette\atversim<}}
\def\gsim{\mathrel{\mathpalette\atversim>}}
\def\sigv{\ifmmode \langle\sigma v\rangle\else $\langle\sigma v\rangle$\fi}
\newskip\zatskip \zatskip=0pt plus0pt minus0pt
\def\matth{\mathsurround=0pt}
\def\lsim{\mathrel{\mathpalette\atversim<}}
\def\gsim{\mathrel{\mathpalette\atversim>}}
\def\atversim#1#2{\lower0.7ex\vbox{\baselineskip\zatskip\lineskip\zatskip
  \lineskiplimit
  0pt\ialign{$\matth#1\hfil##\hfil$\crcr#2\crcr\sim\crcr}}}

\usepackage{graphicx,type1cm,eso-pic,color}


\begin{document}
\preprint{UCI--TR--2016--14}
\thispagestyle{empty}
\vspace*{-3.5cm}

\vspace{0.5in}

\title{ Searching for spin-3/2 leptons}

\begin{center}
\begin{abstract}
A model of spin-3/2 vector-like leptons is introduced, which serve as an alternative probe of possible lepton compositeness.  Possible signatures at the Large Hadron Collider are explored, and current data are used to constrain the masses of the hypothetical new leptons.
\end{abstract}
\end{center}

\author{Mohammad Abdullah}
\author{Kevin Bauer}
\author{Luis Gutierrez}
\author{John Sandy}
\author{Daniel Whiteson}
\affiliation{Department of Physics and Astronomy, University of
  California, Irvine, CA 92697}

\maketitle


\section{Introduction}

A major open question in modern particle physics is whether the
standard model fermions are fundamental particles, or whether they are
composite, such as stable states of more fundamental particles~\cite{Eichten:1983hw}\cite{Contino:2006nn}\cite{Lillie:2007hd}\cite{Chivukula:1987py}\cite{Shupe:1979fv}.  

Such compositeness could be identified by the observation of excited states, which has been studied in depth~\cite{Dicus:2012uh}.  Here, we propose an alternative possibility, the existence of vector-like spin-3/2 copies of the standard model fermions, in which the left- and right-handed fermions have identical charges.

In this paper, we propose such a model, discuss the potential signals at the Large Hadron Collider (LHC) and apply existing LHC data to calculate constraints on the mass of spin-3/2 leptons.

\section{Theory}

We assume that the new spin-3/2 lepton fields are charged identically
to the standard model leptons but are vector-like; specifically,  the
right- and left-handed fields couple identically, for otherwise the
model would be heavily constrained by a combination of constraints
from electroweak precision data and Higgs boson observables \cite{Lenz:2013iha}. This still leaves two options: the leptons can be $SU(2)$ doublets, which means they will couple to all electroweak bosons, or $SU(2)$ singlets, which means they will only couple to photons and $Z$ bosons. The doublet model will be the focus of this work due to its richer structure; we will supplement it with effective operators that allow for decays. 

The Lagrangian for such a model can be written as:

\begin{equation}
\mathcal{L} = \mathcal{L}_{\textrm{free}} + \mathcal{L}_{\textrm{gauge}}+ \mathcal{L}_{\textrm{vector}}
\end{equation}

The first two terms contain the mass and kinetic terms and are given by

\begin{equation}
\mathcal{L}_{\textrm{free}} + \mathcal{L}_{\textrm{gauge}}  =  \frac{1}{2}\epsilon^{\mu\nu\rho\sigma}\bar{L}^{*}_{\mu}\gamma_{5}\gamma_{\sigma}D_{\nu}L^{*}_{\rho} + im_{L}\bar{L}^{*}_{\mu}\gamma^{\mu \nu}L^{*}_{\nu}
\end{equation}

\noindent
where $D$ is the covariant derivative, $\ell$ can be either $e$, $\mu$, or $\tau$ and the $*$ denotes the spin-3/2 leptons. If we expand the derivative term, the free part of the Lagrangian will be

\begin{eqnarray}
\mathcal{L}_{\textrm{free}} = \frac{1}{2}\epsilon^{\mu\nu\rho\sigma}(\ell^{*+}_{\mu}\gamma_{5}\gamma_{\sigma}\partial_{\nu}\ell^{*-}_{\rho}+\bar{\nu}^{*}_{\ell\mu}\gamma_{5}\gamma_{\sigma}\partial_{\nu}\nu^{*}_{\ell\rho}) \nonumber \\ 
+ im_{L}\ell^{*+}_{\mu}\gamma^{\mu \nu}\ell^{*-}_{\nu} + im_{L}\bar{\nu}^{*}_{\ell\mu}\gamma^{\mu \nu}\nu^{*}_{\ell\nu}
\end{eqnarray}

Note that the charged and neutral components have the same mass ($m_{L}$) due to $SU(2)$ gauge invariance. Expanding the covariant derivative we get the interactions with the standard model vector bosons which, after electroweak symmetry breaking, look like

\begin{eqnarray}
\mathcal{L}_{\textrm{gauge}} &=& \epsilon^{\mu \nu \rho \sigma}[2ie_{W} \left(W^{+}_{\nu}\bar{\nu}^{*}_{\ell\mu}\gamma_{5}\gamma_{\sigma}\ell^{*-}_{\rho}+W^{-}_{\nu}\ell^{*+}_{\mu}\gamma_{5}\gamma_{\sigma}\nu^{*}_{\ell\rho}\right) \nonumber \\
&+&  \frac{ie(-\frac{1}{2}+s_{W}^{2})}{s_{W}c_{W}}Z_{\nu}\ell^{*+}_{\mu}\gamma_{5}\gamma_{\sigma}\ell^{*-}_{\rho}  \nonumber \\
&+& \frac{ie}{2s_{W}c_{W}}Z_{\nu}\bar{\nu}^{*}_{\ell\mu}\gamma_{5}\gamma_{\sigma}\nu^{*}_{l\rho} \nonumber \\
&+& -ieA_{\nu}\ell^{*+}_{\mu}\gamma_{5}\gamma_{\sigma}\ell^{*-}_{\rho}]
\end{eqnarray}

\noindent
where all couplings are well measured electroweak parameters given approximately by 

\begin{eqnarray*}
e &=& \sqrt{4\pi \alpha} \simeq 0.3\\
e_{W} &\equiv& \frac{e}{s_{W}2\sqrt{2}}\simeq 0.2 \\
s_{W} &\equiv&  sin\theta_{W}  \simeq 0.5 \\
c_{W} &\equiv&  cos\theta_{W}  \simeq 0.9
\end{eqnarray*}

Next, we move to the effective operators without which the new fields
would be stable; note that the charged and neutral components of the new leptons have identical masses. Since the compositeness scale is higher than the electroweak scale, it makes more sense to write down operators that respect the underlying $SU(2)$ symmetry of the theory which reduces the number of allowed terms and parameters.  For concreteness and simplicity we will assume only the following effective operators

\begin{eqnarray}
\mathcal{L}_{\textrm{vector}} &=& -\frac{i}{2}\frac{1}{\Lambda_{B}}B^{\mu\nu} (\bar{L}^{*}_{\mu}\gamma_{\nu}L + \bar{L}\gamma_{\nu}L^{*}_{\mu}) \nonumber \\
&+& \frac{i}{2}\frac{1}{\Lambda_{W}}W^{\mu\nu a} (\bar{L}^{*}_{\mu}\sigma_{a}\gamma_{\nu}L + \bar{L}\sigma_{a}\gamma_{\nu}L^{*}_{\mu})\nonumber \\
&\equiv&-\frac{i}{2}g_{B}B^{\mu\nu} (\bar{L}^{*}_{\mu}\gamma_{\nu}L + \bar{L}\gamma_{\nu}L^{*}_{\mu}) \nonumber \\
&+& \frac{i}{2}g_{W}W^{\mu\nu a} (\bar{L}^{*}_{\mu}\sigma_{a}\gamma_{\nu}L + \bar{L}\sigma_{a}\gamma_{\nu}L^{*}_{\mu})
\end{eqnarray}

\noindent
where $\Lambda_{B}$ and $\Lambda_{W}$ are scales generated by some UV physics to which we are oblivious. 

Other operators that we could have considered are 4-fermion operators
involving both new and standard model fields, in which case the final
state leptons are direct decay products of the spin-3/2 fields rather
than secondary particles making them easier to detect. Alternatively
we could have chosen  electroweak symmetry breaking operators that
lead to mass splitting between the neutral and charged fields allowing
one to decay to the other. We leave an exploration of such possibilities
for future work.

\section{Bounds}
Since the new fields are vector-like and do not couple to the Higgs at tree level, their electroweak charge causes no conflict with measurement of $S,T,U$ parameters~\cite{ALEPH:2005ab}\cite{Peskin:1990zt} (or, effectively, the electroweak bosons' self-energy). The effective operators are, in principle, constrained by measurements of the anomalous magnetic moment of charged leptons, especially that of the muon~\cite{PhysRevD.86.010001}. Such constraints can easily be avoided if the couplings $g_{B}$ and $g_{W}$ are sufficiently small. Ensuring that these couplings are small also ensures that the effective operators do not compete with the gauge interactions in the production of the new fermions. On the other hand, if the couplings are too small the new fields would correspond to particles with long lifetimes and would be severely constrained by long-lived charged particle searches \cite{Chatrchyan:2013oca}\cite{ATLAS:2014fka}\cite{Aad:2015qfa}. The range $10^{-9}\;\text{GeV}^{-1}<g_{B},\;g_{W}<10^{-4}\;\text{GeV}^{-1}$ satisfies these requirements and we will assume such values for the remainder of the work. We emphasize that the exact values are, for a given ratio of $g_{B}$ and $g_{W}$, unimportant since they modify neither the branching ratios nor the production rate.

Collider bounds on spin-1/2 vector-like leptons are potentially applicable. However, due to different choices in either the production or decay channels, none of the analyses in the literature are directly applicable. Moreover, it is possible that the difference in kinematics due to the higher spin are significant enough to alter the limit, so it is worthwhile to study this particular case in addition.

\section{Collider Phenomenology}

The production of the new fields is predominantly through Drell-Yan pair production with electroweak gauge bosons as mediators, and the cross sections are fixed by standard model couplings for a given mass. As explained earlier, the couplings of the effective operators are chosen to be small enough to contribute sub-dominantly to production even at the highest mass considered.  

\begin{figure}[h]
\centering
\vspace{0.5cm}
\includegraphics[width=0.5\textwidth]{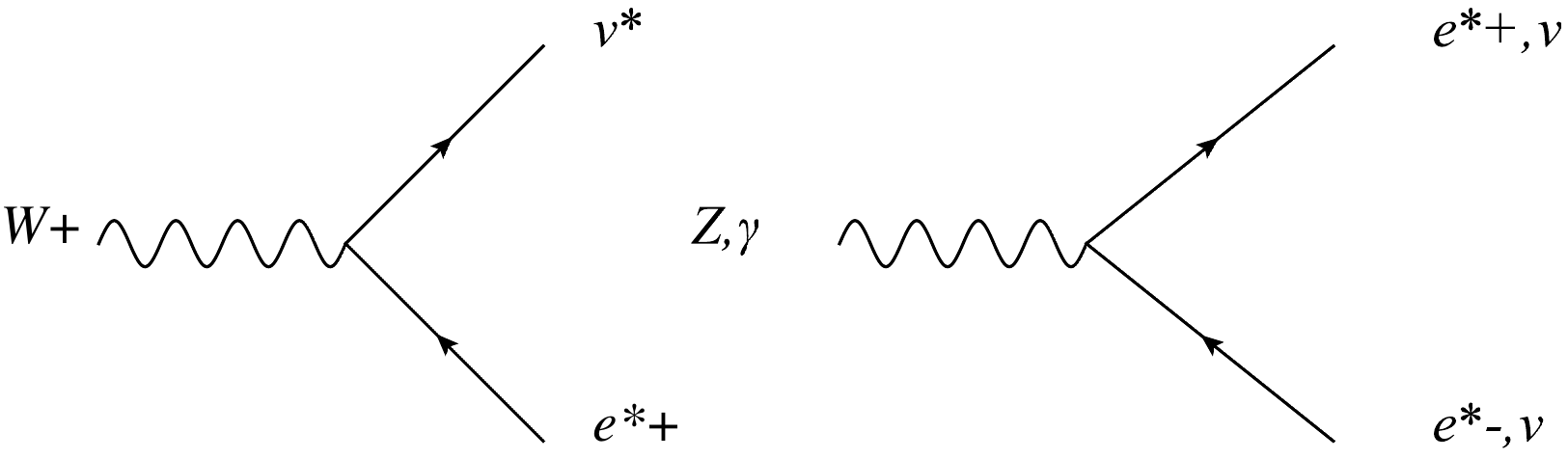}
\caption{Production modes of the spin-3/2 leptons via standard model electroweak bosons.}
\label{fig:prod}
\end{figure}

The pairs that can be produced are $l^{*+}l^{*-}$, $\bar{\nu}^{*}_{l}\nu^{*}_{l}$, $l^{*+}\nu^{*}_{l}$, and $l^{*-}\bar{\nu}^{*}_{l}$, with the last two being the most common since $W$ production at the LHC is the larger of the three bosons. As for decays, the effective operators open up a variety of channels. The allowed decays are shown in the table below. 

\begin{table}[h]
\caption{Allowed decay modes for spin-3/2 leptons $\ell^*$ and $\nu^*$.}
\begin{center}
$
\begin{array}{c | c c c c c c}
\ell^{*-}	&	Z \ell^{-}	&	\gamma \ell^{-}	&	W^{-}\nu_{\ell}	&	W^{-}Z\nu_{\ell}	&W^{-}\gamma \nu_{\ell}	&W^{+}W^{-}\ell^{-}\\
\nu^{*}_{\ell}	&	Z \nu_{\ell}	&	\gamma \nu_{\ell}	&	W^{+}\ell^{-}	&	W^{+}Z\ell^{-}	&W^{+}\gamma \ell^{-}	&W^{+}W^{-}\nu_{\ell}
\end{array}
$
\label{tab:decays}
\end{center}
\end{table}

By contrived choices of $g_{W}$ and $g_{B}$ we can turn off one of the decays in the first two columns. All the other columns depend only on $g_{W}$ and can only be turned off by setting it to zero, in which case the decays in the first two columns must either be all on or all off. One good feature of this set of decays is that there will always be some visible decay products, even if we produce a pair of spin-3/2 neutrinos. 

In this work will limit ourselves to 2-body decays including electron and muons, which results in promising signals; we leave the photon channels for a future work. We will place the limits on the  benchmark points listed in Table~\ref{points} where the forbidden decay channel is indicated in each case.

\begin{table}[h]
\begin{center}
\begin{tabular}{c | c c c c}
Benchmark Point 	&	$g_{W}$ & :	& $g_{B}$ 	& Forbidden Channels\\
\hline
1	&	1	&	:& 1		&None				\\
2	&	$c_{W}$&	:	& $s_{W}$		&$\nu^*\rightarrow \gamma \nu$				\\
3	&	$c_{W}$&	:	&-$s_{W}$		&$e^*\rightarrow \gamma e$				\\
4	&	$s_{W}$&	:	&$c_{W}$		&$e^*\rightarrow Z \nu$				\\
5	&	$s_{W}$&	:	&-$c_{W}$		&$\nu^*\rightarrow Z \nu$				\\
6	&	0	&	:&1		&All decays including W's 				
\end{tabular}
\caption{The ratios of $g_{W}$ and $g_{B}$ and the main decay feature of each benchmark point}
\label{points}
\end{center}
\end{table}

\section{Limits from LHC data}

Assuming production via an intermediate electroweak boson and the decays given in Table~\ref{tab:decays}, there are ten distinct production modes for the LHC:

\[ W\rightarrow\ell^*\nu^* \rightarrow Z\ell Z\nu, W\nu Z\nu,W\nu W\ell,Z\ell W\ell \]
\[ Z\rightarrow\ell^*\ell^* \rightarrow Z\ell Z\ell, W\nu W\nu,Z\ell W\nu \]
\[ Z\rightarrow\nu^*\nu^* \rightarrow Z\nu Z\nu, Z\nu W\ell,W\ell W\nu \]

\noindent
which generate final states including many charged leptons, which offer low background rates at the LHC.

We apply selections from the same-sign $2\ell$~\cite{ATLAS:2014kca} and $3\ell$~\cite{Aad:2014hja} searches at ATLAS.   We provide a brief summary below of the signal regions considered.

\begin{figure}
\vspace{0.5cm}
\includegraphics[scale =0.38]{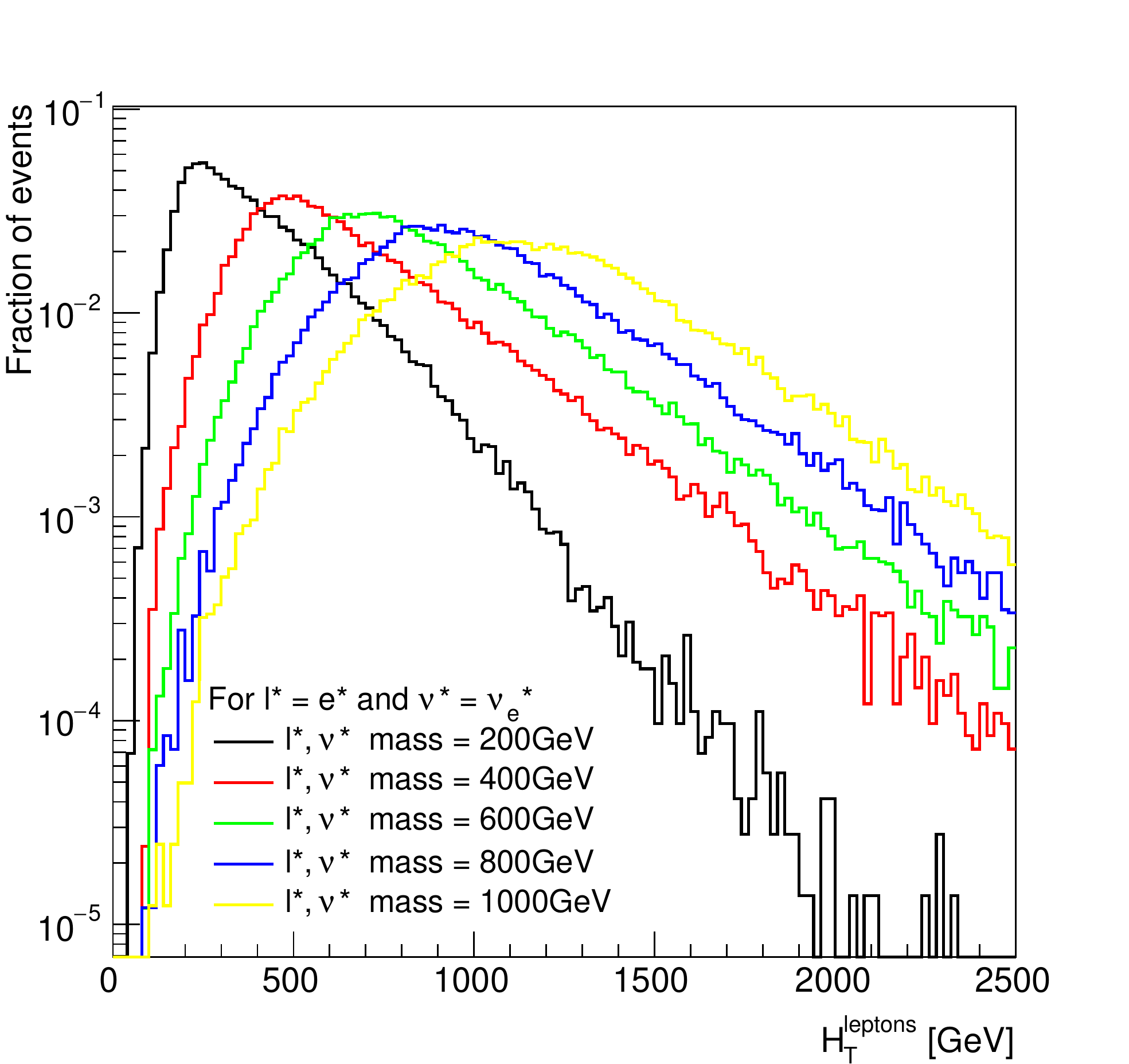}
\includegraphics[scale =0.38]{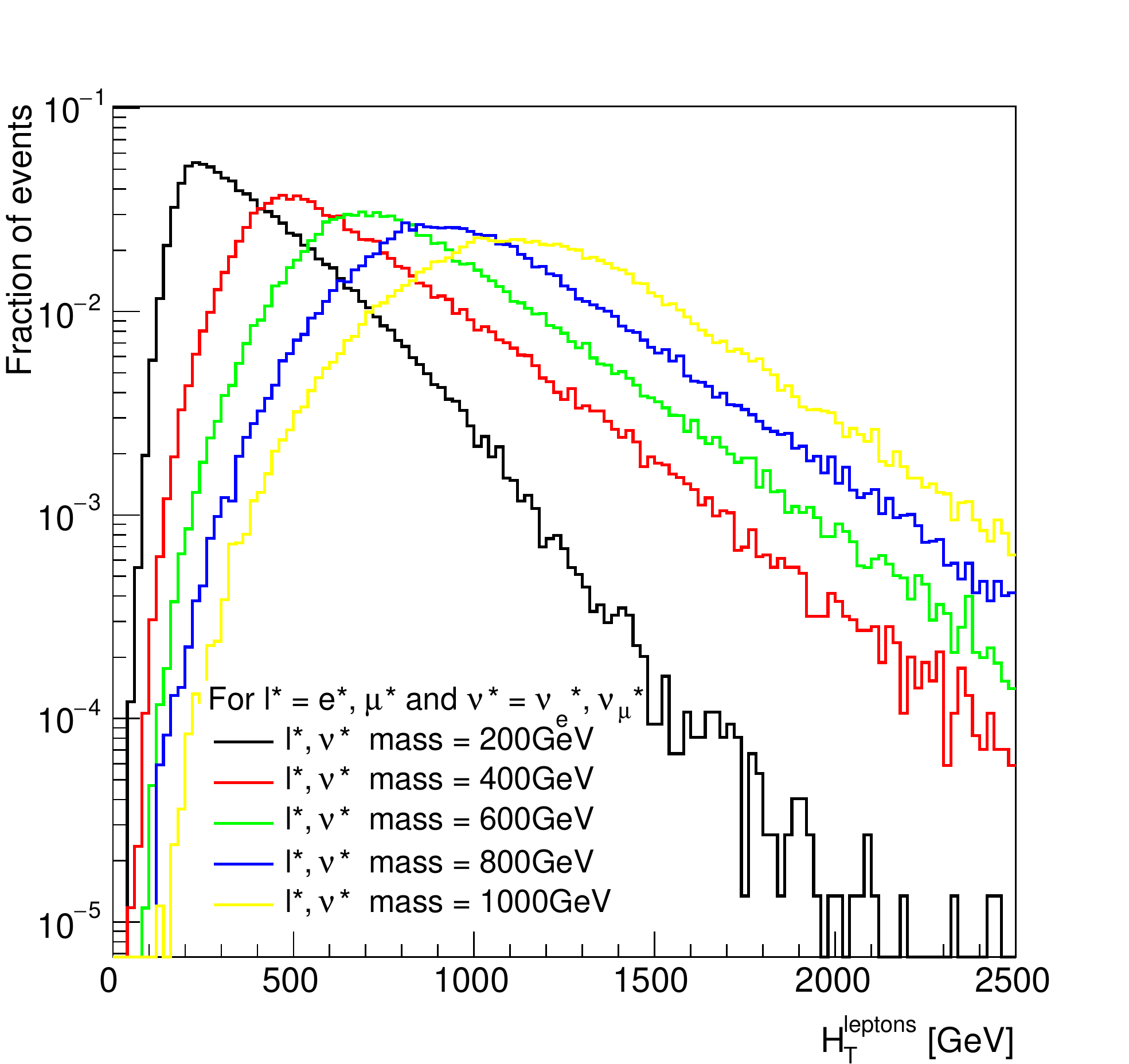}
\caption{Distribution of $H_{\textrm{T}}^{\textrm{leptons}}$ for simulated spin-3/2 lepton samples; the selection required $H_{\textrm{T}}^{\textrm{leptons}}>500$ GeV.  The top pane shows the case where only electrons are considered; the bottom pane shows the case where electrons and muons are considered.}
\label{fig:kin}
\end{figure}

For the $2\ell$ search, the final states $e^{\pm}e^{\pm}, e^{\pm}\mu^{\pm}$ and $\mu^{\pm}\mu^{\pm}$ are considered in  categories defined by the dilepton invariant mass.

For the $3\ell$ search, events are classified using the three highest-$p_{\textrm{T}}$ selected leptons, distinguishing between those that do not contain an opposite-sign same-flavour pair (denoted no-OSSF)  and those that do contain an OSSF pair, which are further subdivided into on-Z and off-Z based on the dilepton and three highest-$p_{\textrm{T}}$ selected leptons invariant mass. For each category, there are several signal regions used to characterize events based on the quantities: 
$H^{\textrm{leptons}}_{\textrm{T}}$, the scalar sum of the transverse momentum $p_{\textrm{T}}$ of the three leptons used to categorize the event; 
$p_{\textrm{T}}^{\ell, \textrm{min}}$, the minimum transverse momentum $p_{\textrm{T}}$ of the three leptons used to categorize the event; 
$E_{\textrm{T}}^{\textrm{miss}}$, the missing transverse momentum;
$m_{\textrm{eff}}$,  the scalar sum of $E_{\textrm{T}}^{\textrm{miss}}$, $H_{\textrm{T}}^{\textrm{jets}}$, and the $p_{\textrm{T}}$ of all identified leptons in the event; 
and $m_{\textrm{T}}^{W}=\sqrt{2p_{\textrm{T}}^{\ell}E_{\textrm{T}}^{\textrm{miss}}(1-\cos(\Delta\phi))}$, defined only for on-$Z$ events. For the “transverse mass”, $p_{\textrm{T}}^{\ell}$ corresponds to the transverse momentum of the highest-$p_{\textrm{T}}$ lepton not associated to a $Z$ boson candidate and $\Delta\phi$ is the azimuthal angle between the highest-$p_{\textrm{T}}$ lepton not associated to a $Z$ boson candidate and the missing transverse momentum $E_{\textrm{T}}^{\textrm{miss}}$.

Simulated samples of spin-3/2 leptons were generated with {\sc madgraph5}~\cite{Alwall:2014hca}\cite{Christensen:2013aua} and the decays were performed through {\sc pythia}~\cite{Sjostrand:2006za} for a variety of $\ell^*$ and $\nu^*$ masses under two scenarios. First, we consider the simple case in which $\ell^*=e^*$ and $\nu^*=\nu_e^*$; second, we extend to the second generation and allow $\ell^*=e^*,\mu^*$, $\nu^*=\nu_e^*,\nu_\mu^*$.  For both, we set $m_{e^{*}}=m_{v_{e}^{*}}=m_{\mu^{*}}=m_{v_{\mu}^{*}}$. We generated samples of production cross sections and decay widths for a range of masses and combined them according to each of the six coupling benchmarks described in Table~\ref{points}.

\begin{figure}
\vspace{0.5cm}
\includegraphics[scale =0.38]{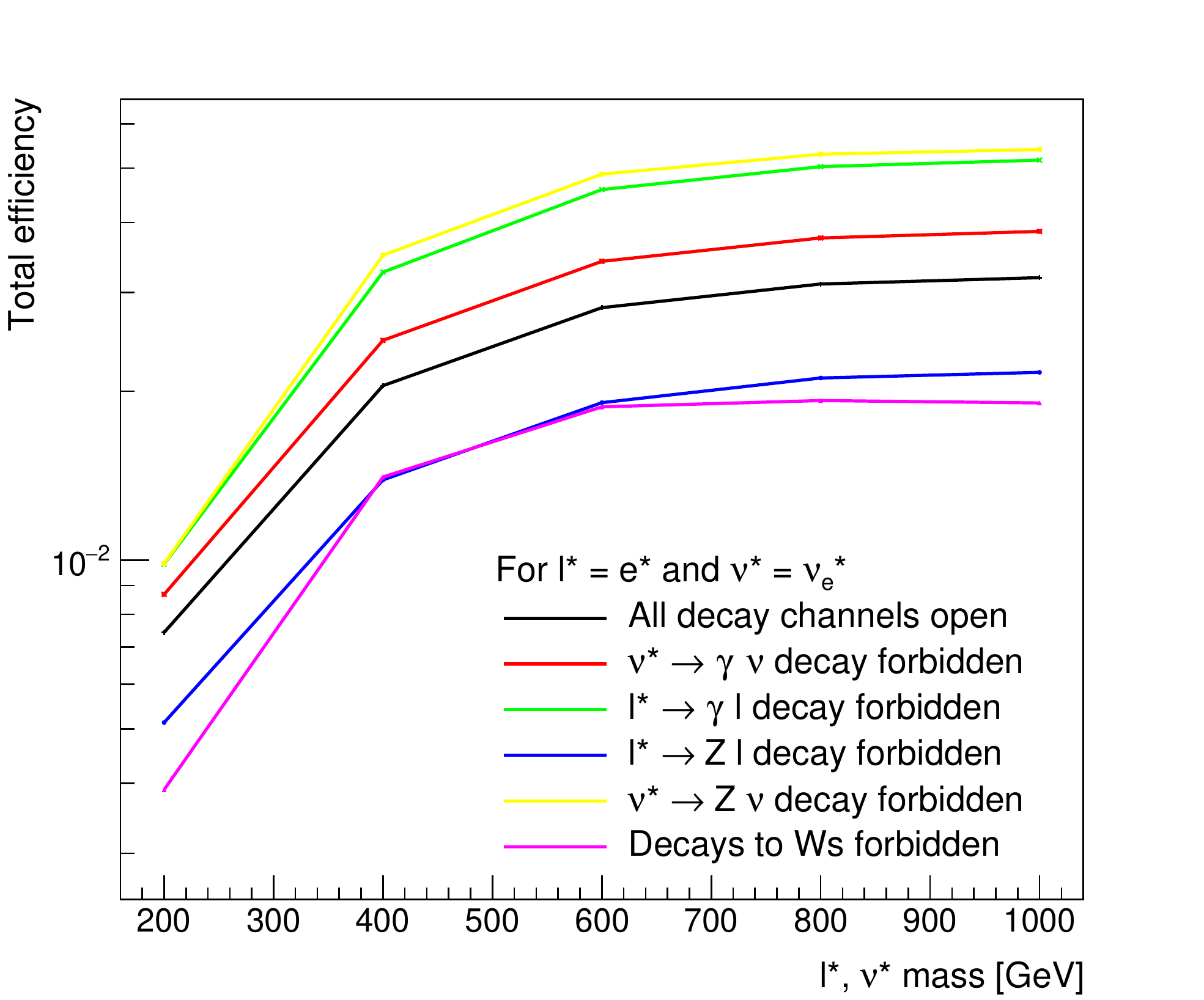}
\includegraphics[scale =0.38]{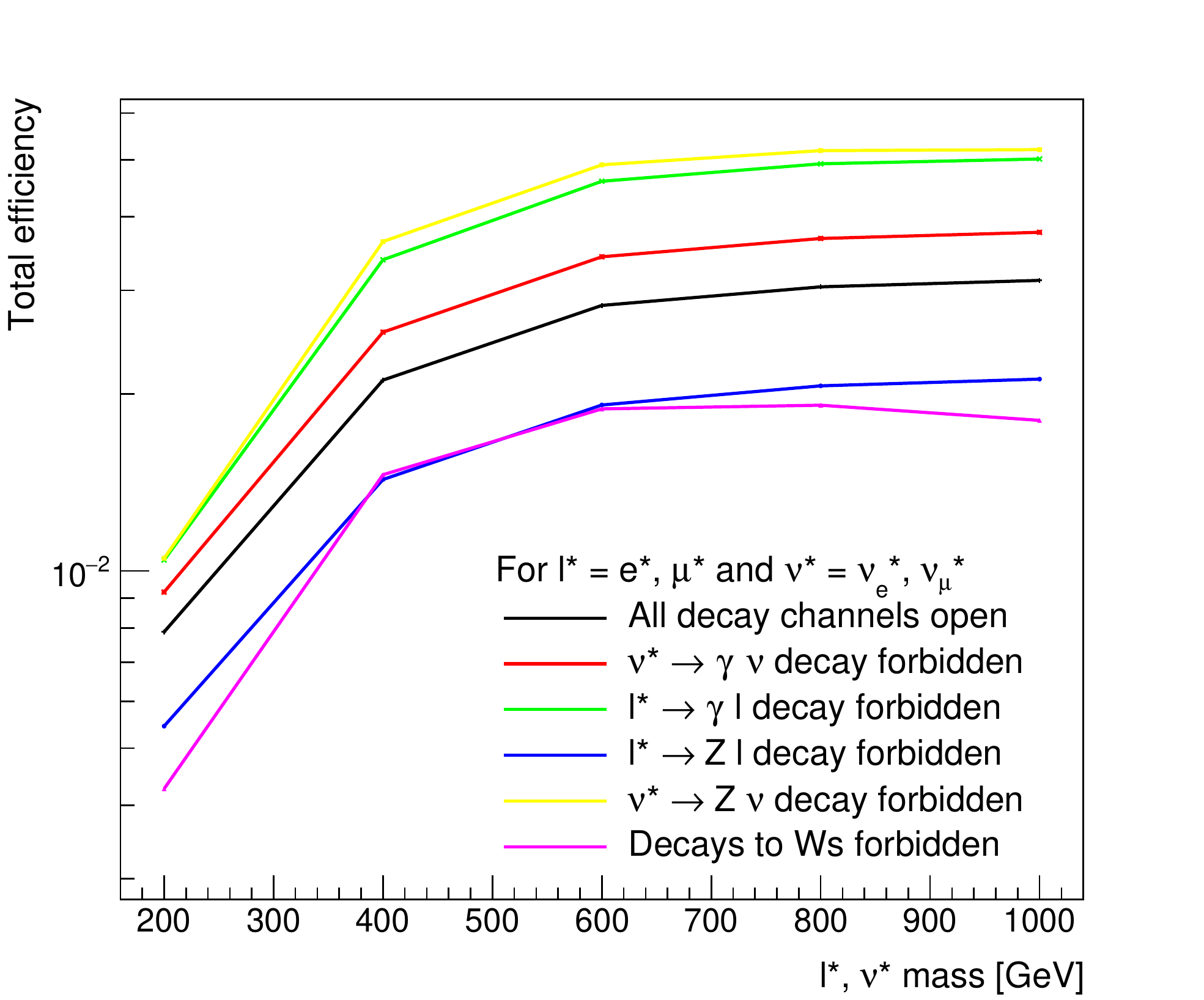}
\caption{Overall efficiency  of the selection defined in Ref.~\cite{Aad:2014hja} which requires $H_{\textrm{T}}^{\textrm{leptons}}>500$ GeV for OSSF off-$Z$, as a function of the spin-3/2 lepton mass. The top pane shows the case where only electrons are considered; the bottom pane shows the case where electrons and muons are considered.}
\label{fig:sel}
\end{figure}

\begin{figure*}
\vspace{0.5cm}
\includegraphics[width=0.4\textwidth]{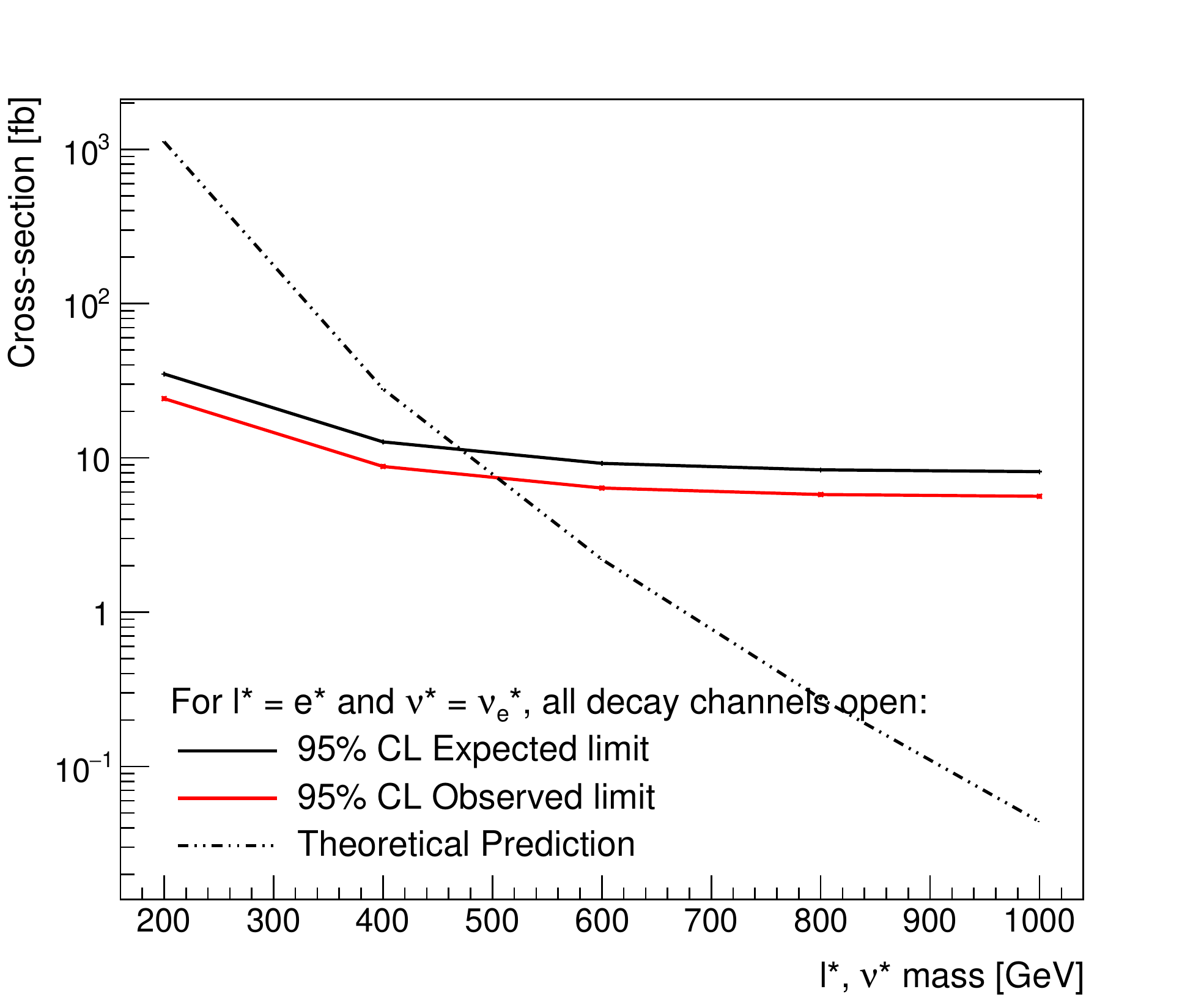}
\includegraphics[width=0.4\textwidth]{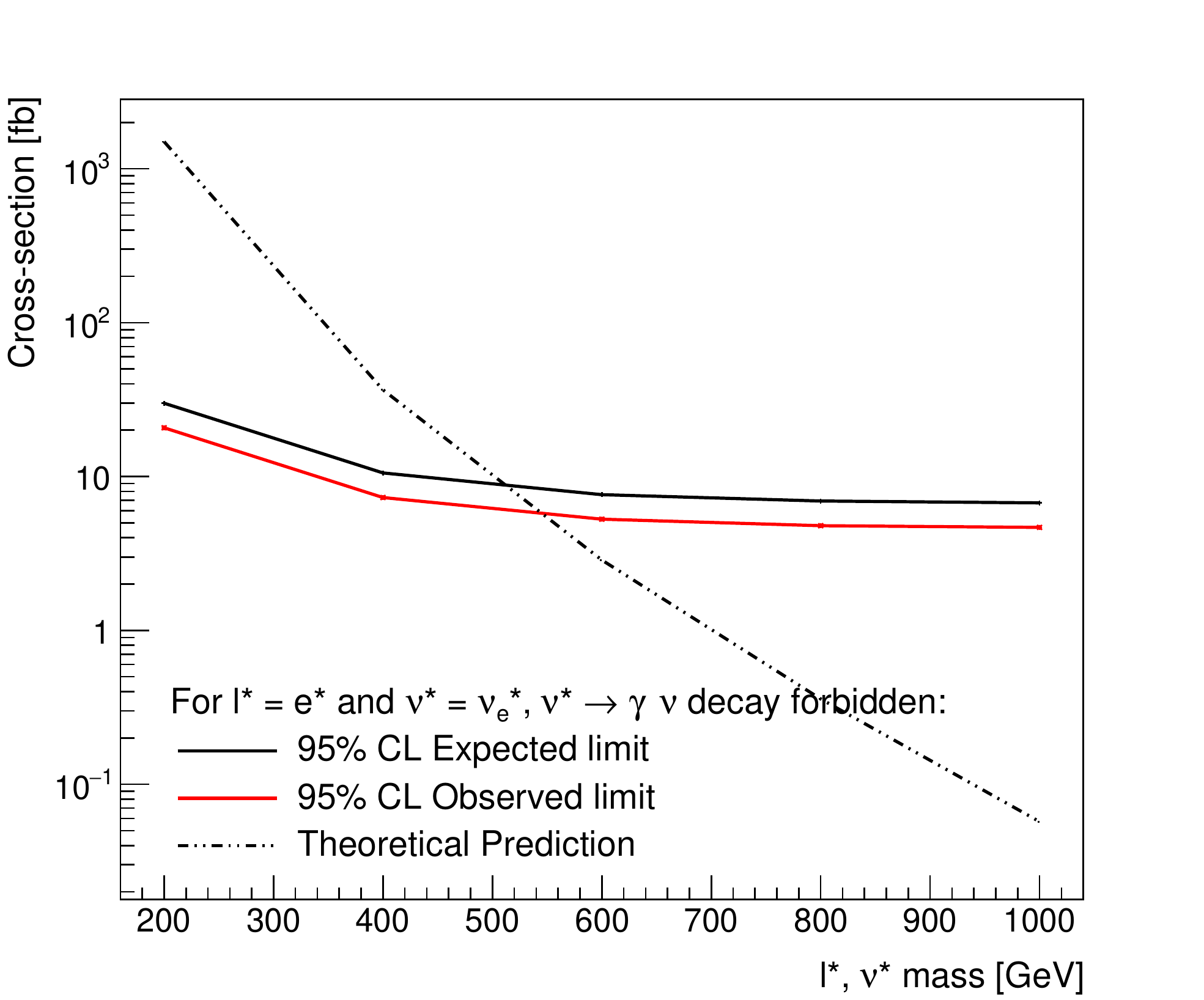}
\includegraphics[width=0.4\textwidth]{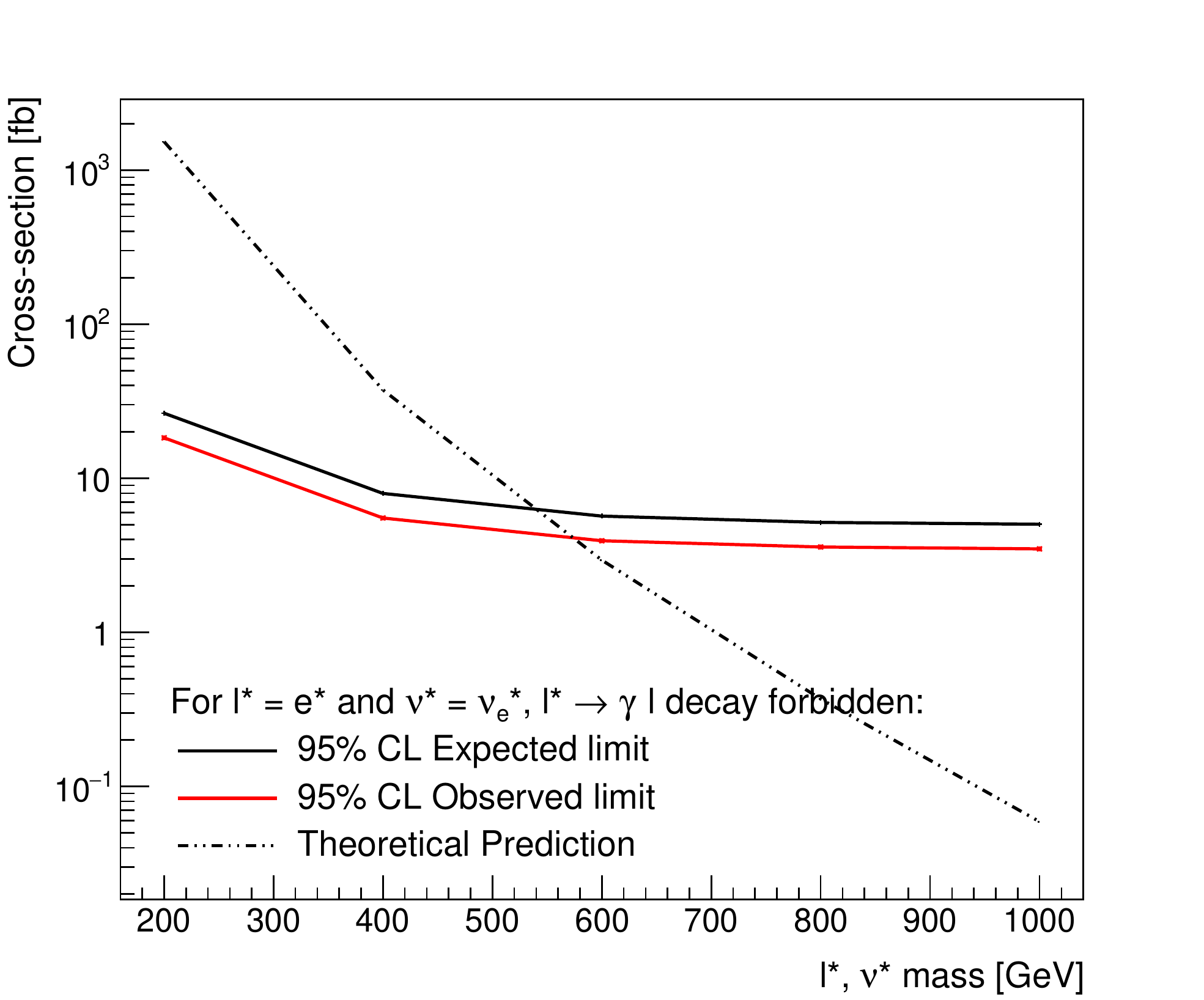}
\includegraphics[width=0.4\textwidth]{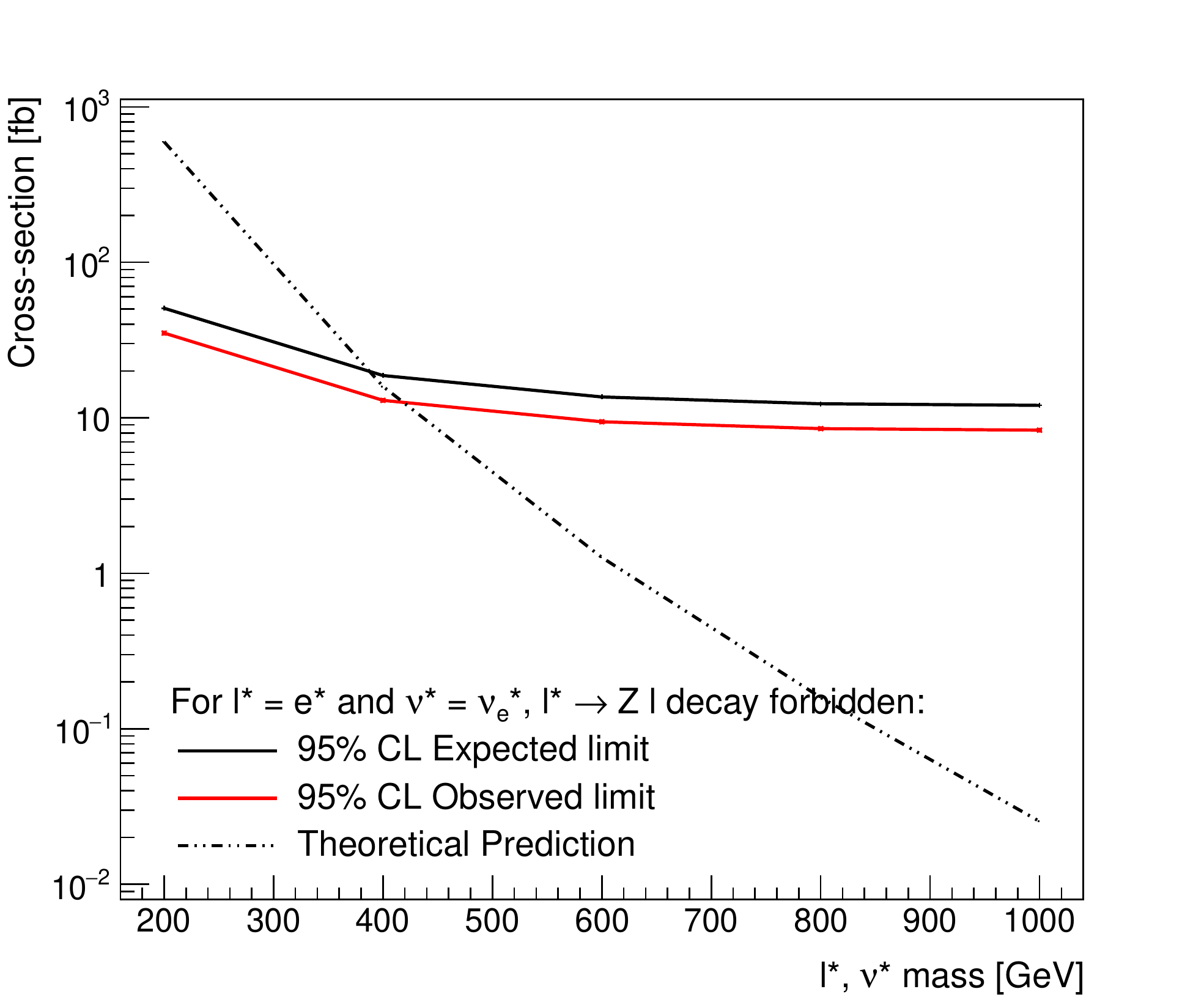}
\includegraphics[width=0.4\textwidth]{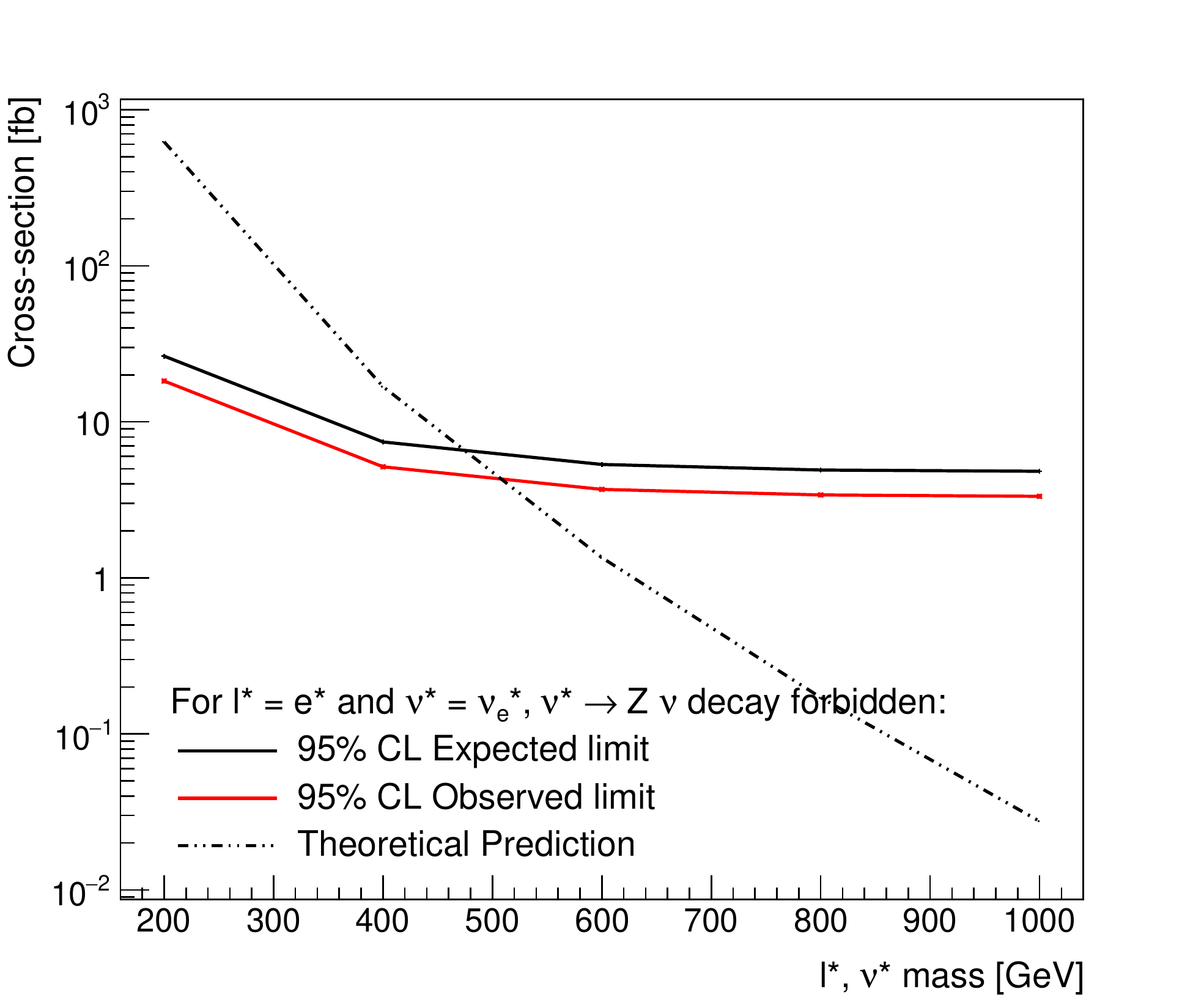}
\includegraphics[width=0.4\textwidth]{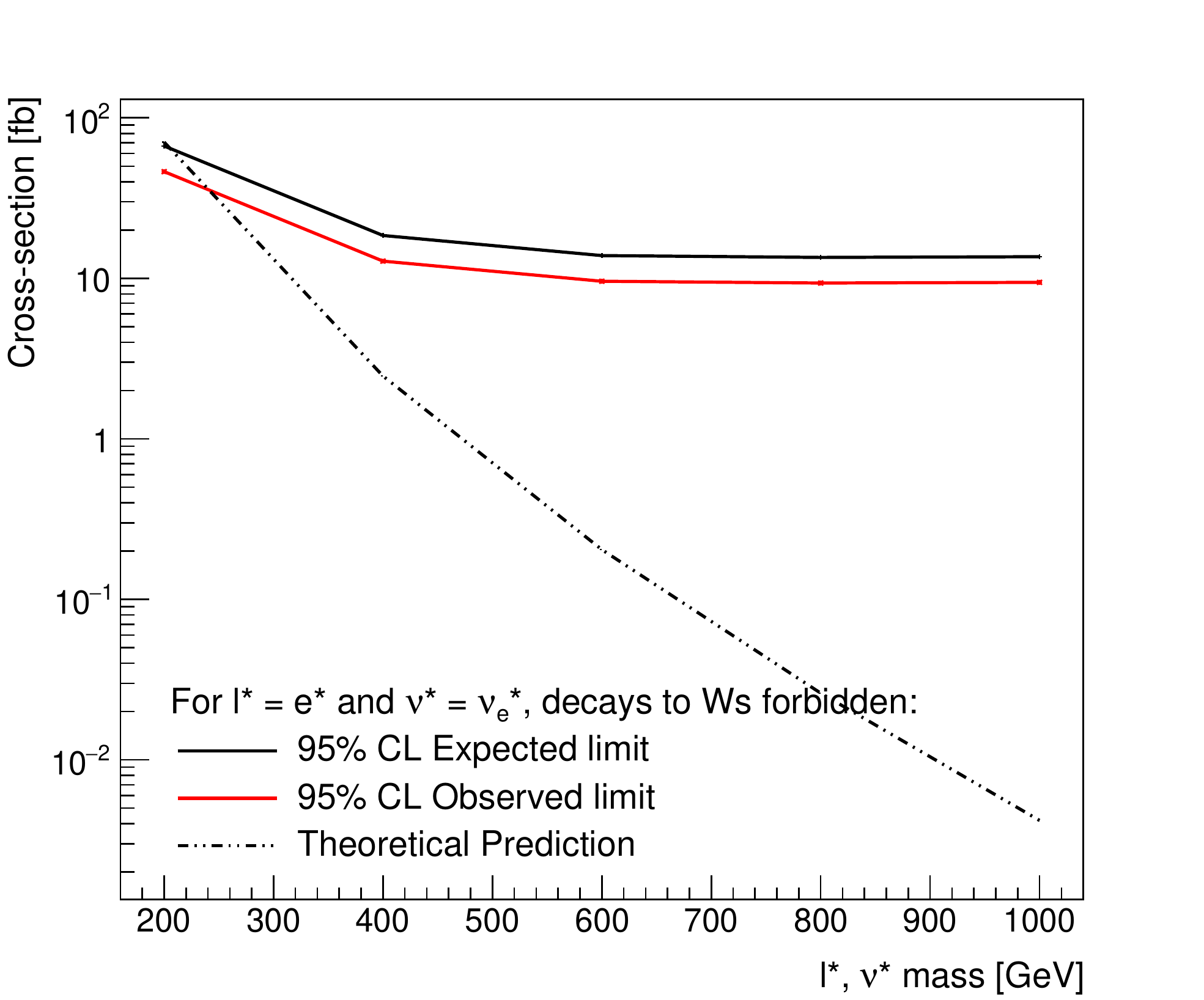}
\caption{Observed and expected limits as a function of $\ell^*$ and $\nu^*$ mass, in the case of $\ell^*=e^*$, $\nu^*=\nu_e^*$
}
\label{fig:lim}
\end{figure*}

\begin{figure*}
\vspace{0.5cm}
\includegraphics[width=0.4\textwidth]{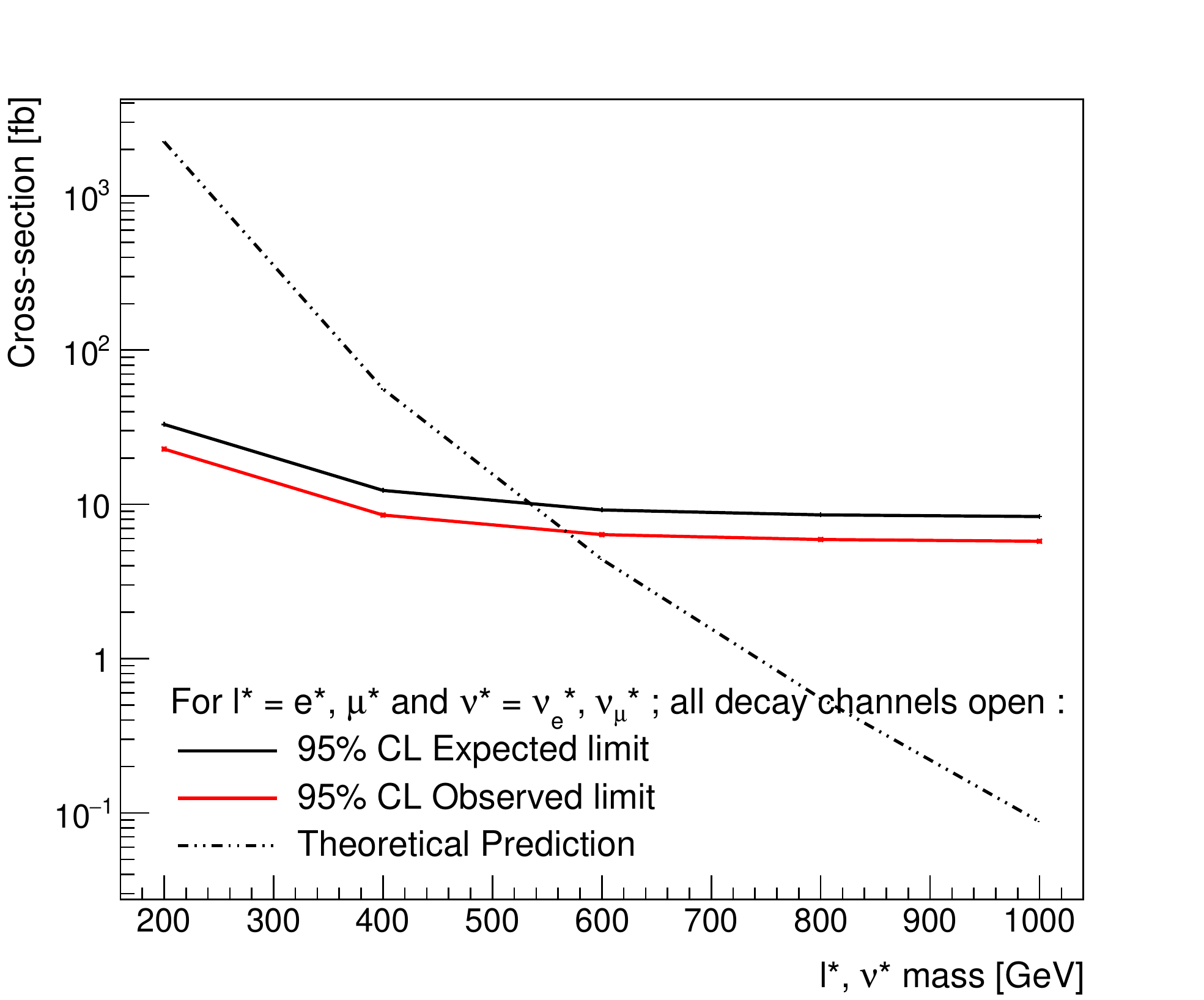}
\includegraphics[width=0.4\textwidth]{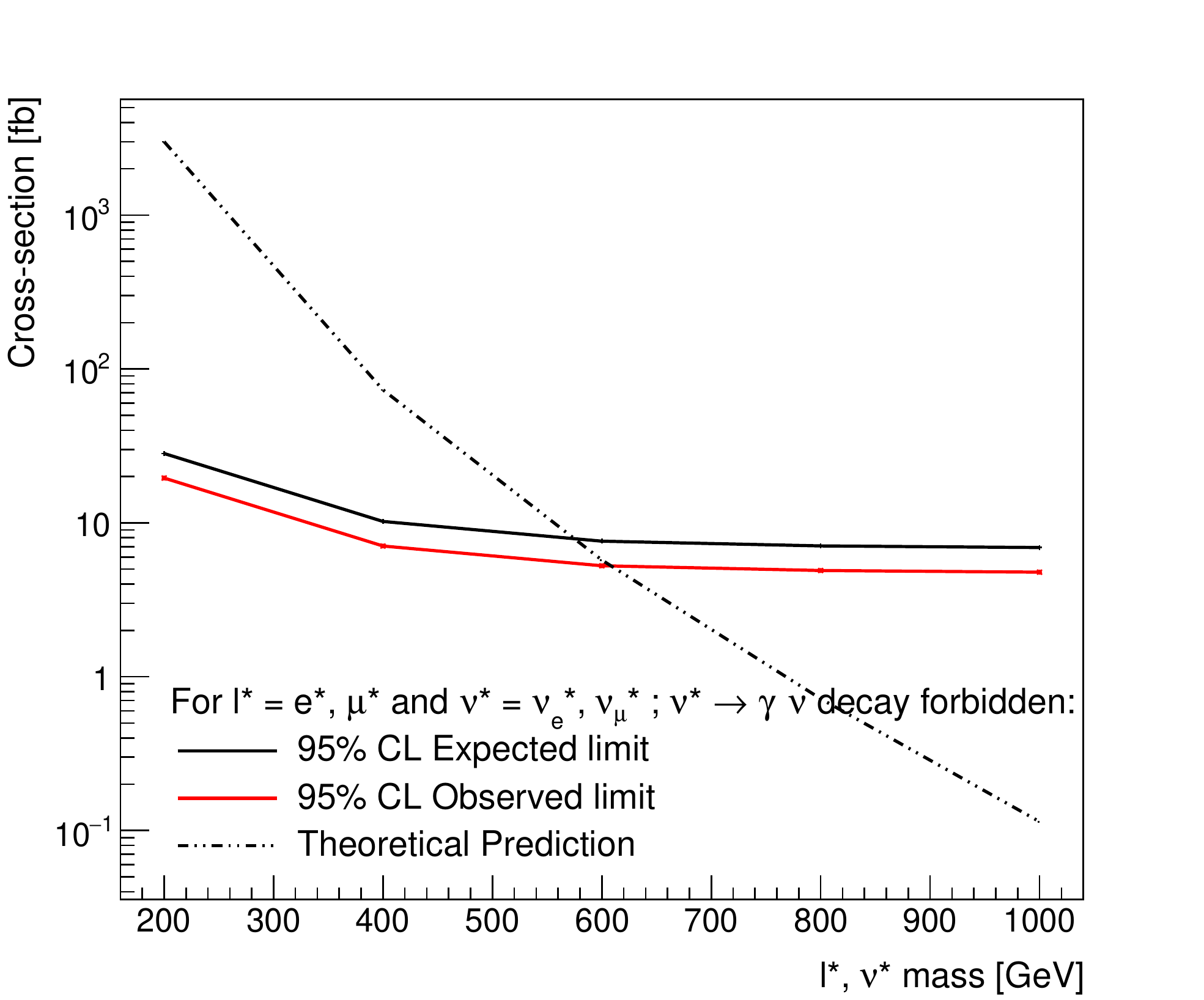}
\includegraphics[width=0.4\textwidth]{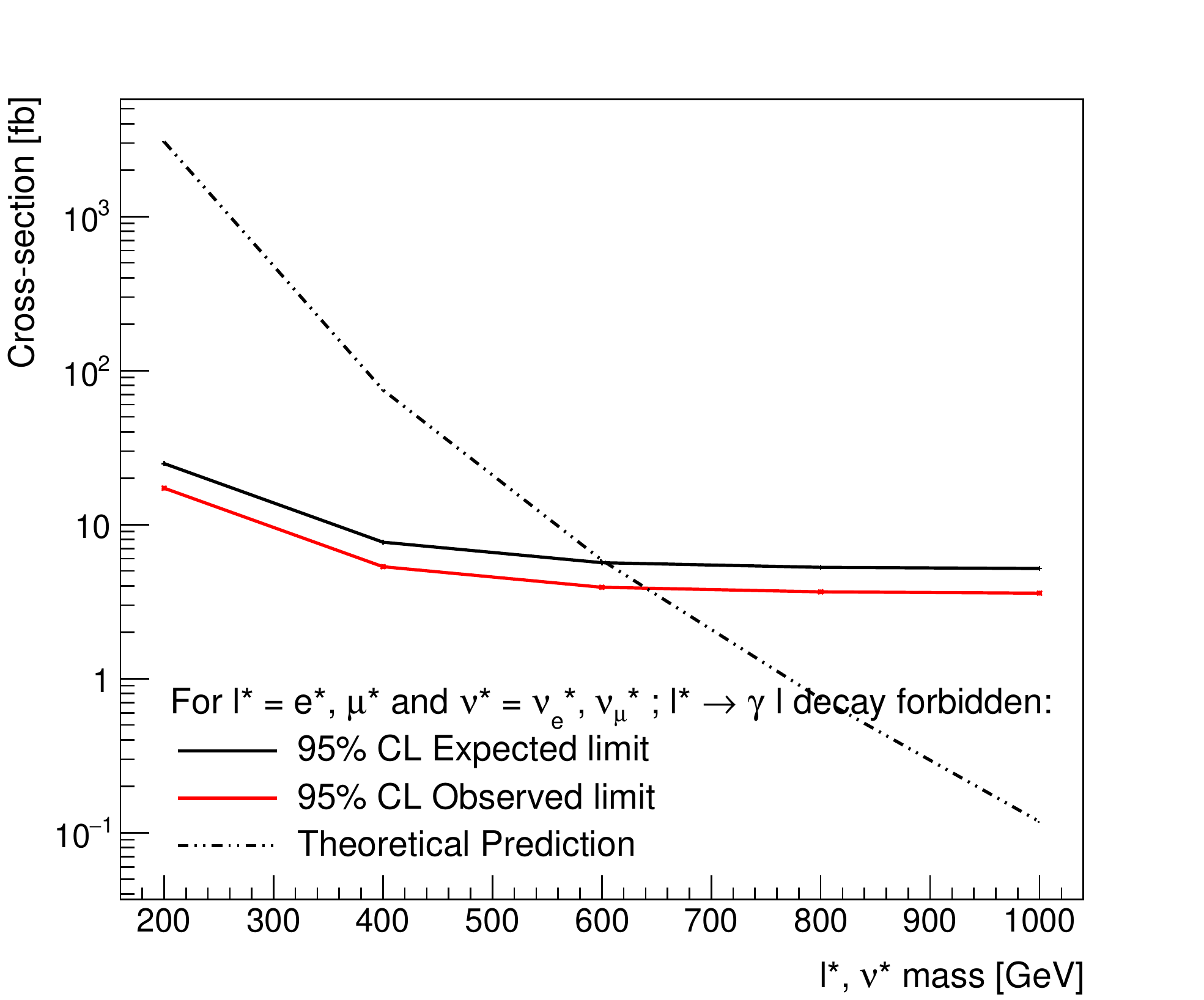}
\includegraphics[width=0.4\textwidth]{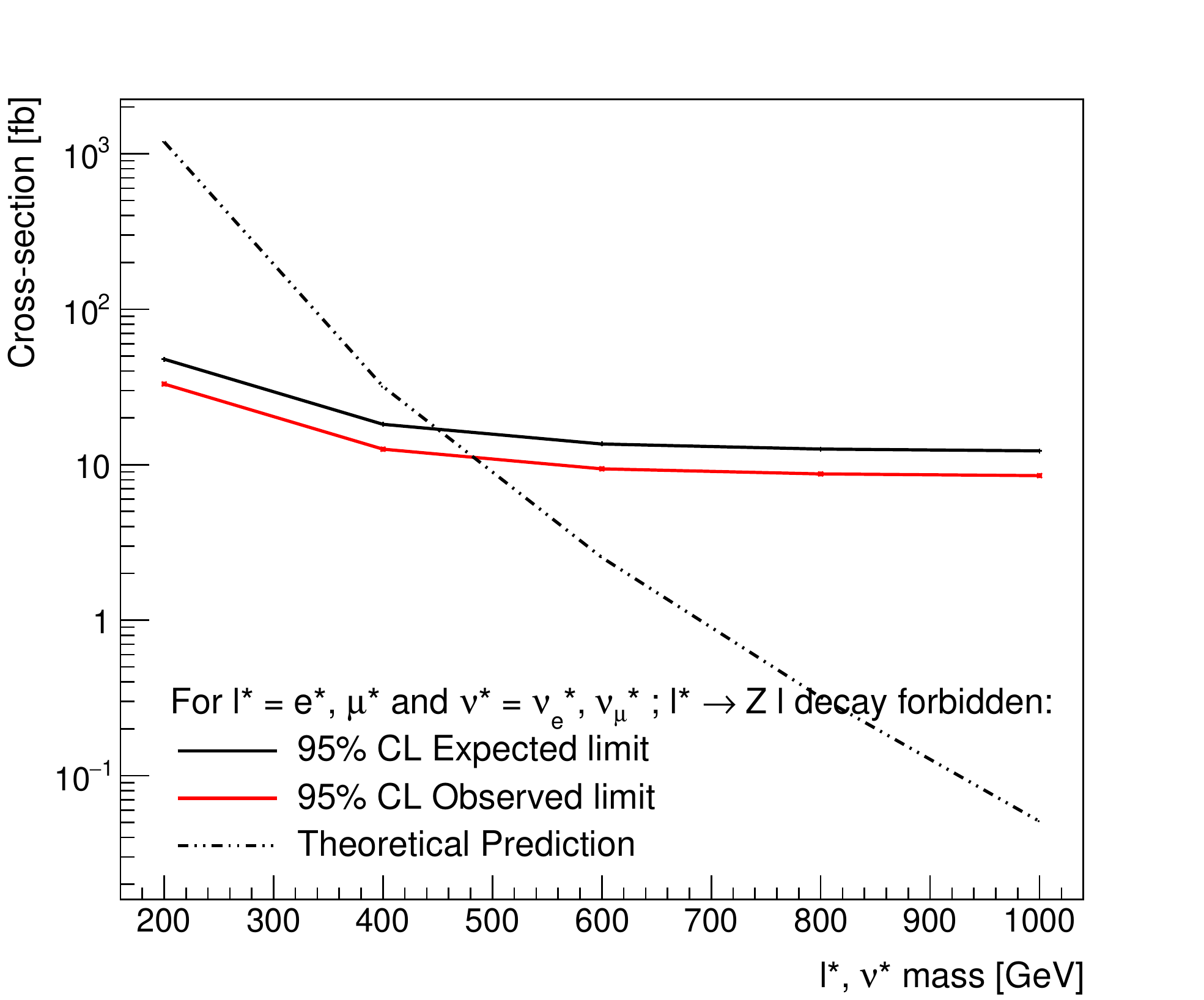}
\includegraphics[width=0.4\textwidth]{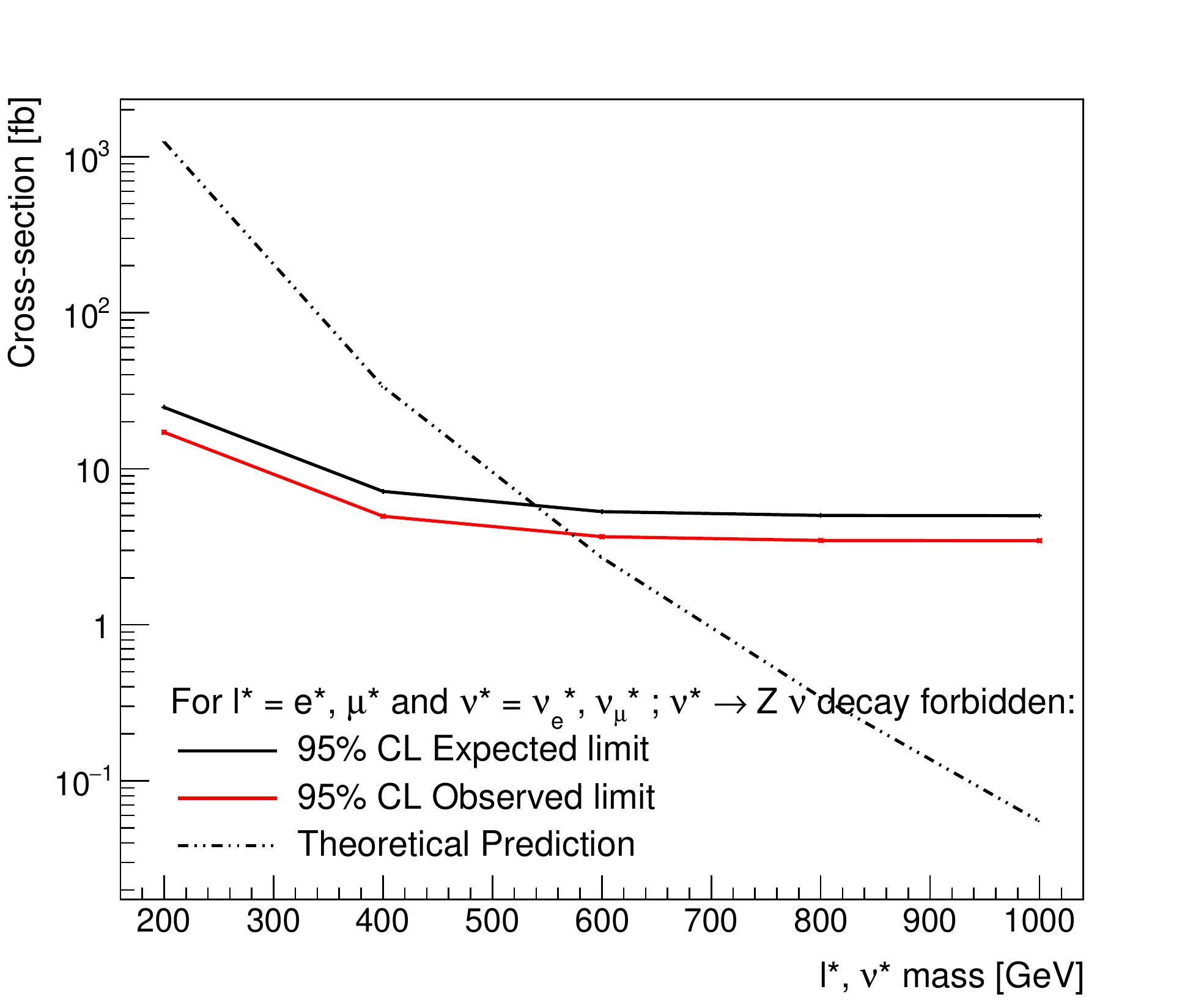}
\includegraphics[width=0.4\textwidth]{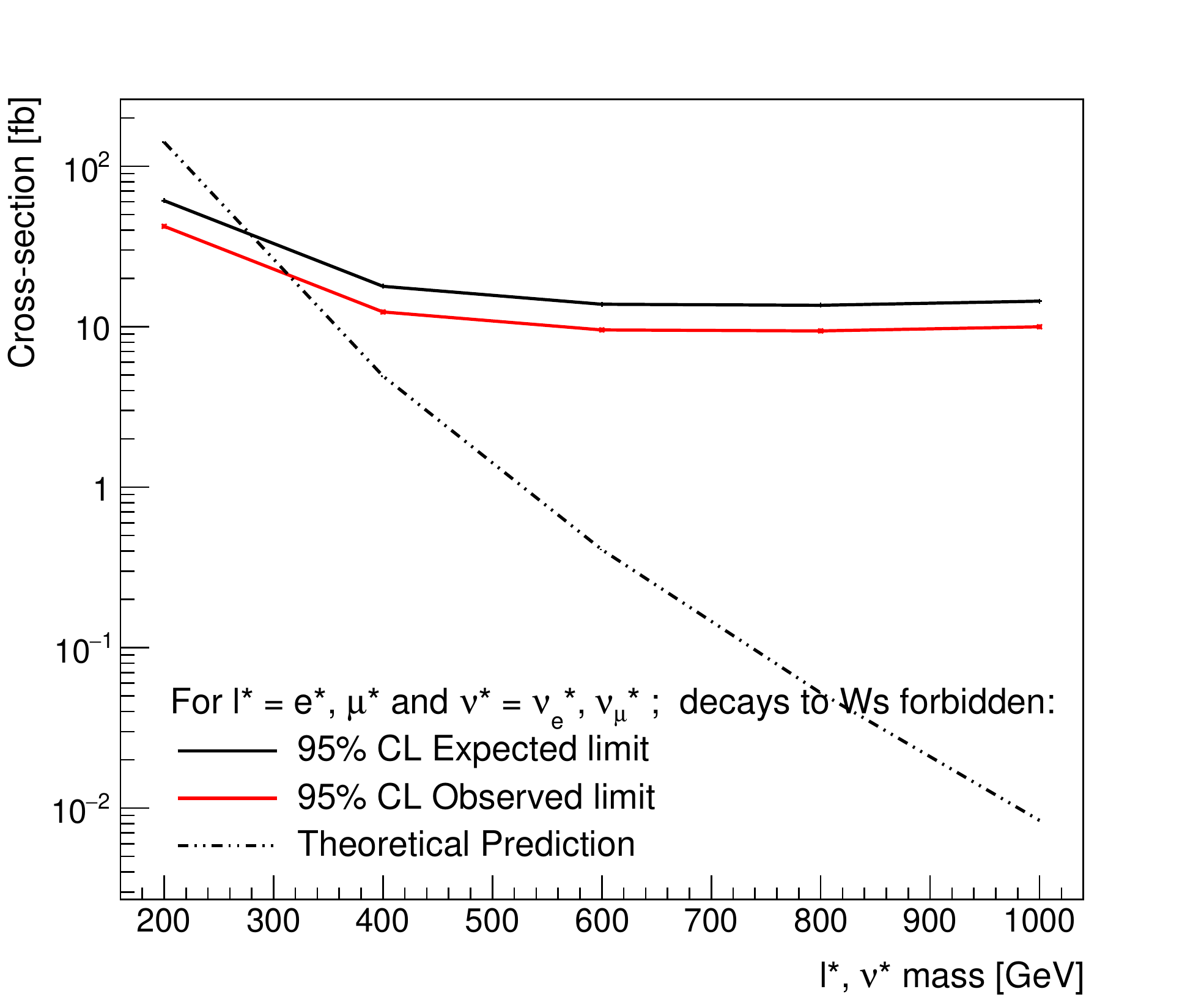}
\caption{Observed and expected limits as a function of $\ell^*$ and $\nu^*$ mass, in the case of $\ell^*=e^*,\mu^*$, $\nu^*=\nu_e^*,\nu_\mu^*$
}
\label{fig:lim2}
\end{figure*}

 From among the signal regions in the two ATLAS searches described
 above, we choose the SR which gives the most powerful expected
 limits: $H_{\textrm{T}}^{\textrm{leptons}}>$ 500 GeV for OSSF off-$Z$.
 Distributions of $H_{\textrm{T}}^{\textrm{leptons}}$ in simulated
 samples after all selection requirements are made are shown in
 Fig.~\ref{fig:kin}.  Selection efficiency is shown in
 Fig.~\ref{fig:sel}. Observed and expected limits are derived
 according to the prescription provided in Ref.~\cite{Aad:2014hja} shown in Fig.~\ref{fig:lim} and Fig.~\ref{fig:lim2}.

\section{Conclusions}
We investigated the sensitivity of existing LHC data to a model of spin-3/2 leptons. Such fields are possible within the context of fermion compositeness which is well studied in the literature. We focused on a scenario where the new fields are dominantly produced by Drell-Yan processes through electroweak bosons and promptly decay through effective operators to electroweak bosons and standard model leptons of the same flavor as the new fields. 

We looked at final states resulting in same-sign $2\ell$ or $3\ell$ and found comparable limits for each. The new fields are ruled out at the 95\% confidence level for masses up to about 560 \gev\, for one new field with electron flavor, and up to about 620 \gev\, for two new, mass degenerate fields, one with electron flavor and one with muon flavor. We expect to get more stringent limits if we include the photon channel.

\section{Acknowledgements}

The authors are grateful to Tim Tait, Bartosz Fornal, and Jordan Smolinsky for helpful
discussion and comments. MA is particularly thankful to Tim Tait for his guidance in constructing the model. This work is supported in part by U.S.~National Science Foundation Grant No.~PHY--1316792.

\bibliographystyle{apsrev}
\bibliography{paper}

\end{document}